\documentclass[lettersize,journal]{IEEEtran}
\usepackage{amsmath,amsfonts}
\usepackage{algorithmic}
\usepackage{algorithm}
\usepackage{array}
\usepackage[caption=false,font=normalsize,labelfont=sf,textfont=sf]{subfig}
\usepackage{textcomp}
\usepackage{stfloats}
\usepackage{url}
\usepackage{verbatim}
\usepackage{graphicx}
\usepackage{cite}
\usepackage{siunitx}
\usepackage{caption} 
\usepackage[pdftex]{hyperref}

\usepackage{xcolor}

\begin{document}
\newcommand{\modelname}{NoiseBandNet}
\newcommand{\fb}{filterbank}

\title{\modelname{}: Controllable Time-Varying Neural Synthesis of Sound Effects Using Filterbanks}

\author{\IEEEauthorblockN{Adri{\'a}n Barahona-R{\'i}os$^{1}$}, \IEEEauthorblockN{Tom Collins$^{2,3}$} \\
\IEEEauthorblockA{$^{1}$Department of Computer Science, University of York, ajbr501@york.ac.uk \\
$^{2}$Frost School of Music, University of Miami\\
$^{3}$Music Artificial Intelligence Algorithms, Inc., tomthecollins@gmail.com}}

\maketitle

\begin{abstract}
    Controllable neural audio synthesis of sound effects is a challenging task due to the potential scarcity and spectro-temporal variance of the data. Differentiable digital signal processing (DDSP) synthesisers have been successfully employed to model and control musical and harmonic signals using relatively limited data and computational resources. Here we propose \modelname{}, an architecture capable of synthesising and controlling sound effects by filtering white noise through a \fb{}, thus going further than previous systems that make assumptions about the harmonic nature of sounds. We evaluate our approach via a series of experiments, modelling footsteps, thunderstorm, pottery, knocking, and metal sound effects. Comparing \modelname{}  audio reconstruction capabilities to four variants of the DDSP-filtered noise synthesiser, \modelname{} scores higher in nine out of ten evaluation categories, establishing a flexible DDSP method for generating time-varying, inharmonic sound effects of arbitrary length with both good time and frequency resolution. Finally, we introduce some potential creative uses of \modelname{}, by generating variations, performing loudness transfer, and by training it on user-defined control curves.
\end{abstract}


{\let\thefootnote\relax\footnotetext{This work has been submitted to the IEEE for possible publication. Copyright may be transferred without notice, after which this version may no longer be accessible.}}

\section{Introduction}

In media, sound effects can be defined as those sound elements other than music or speech \cite{hausman2015modern}. Typical sound effects are, for instance, footsteps or environmental sounds such as rain. This broad definition implies that sound effects may exhibit, within the same category, wide and narrow spectral bands or static and transient amplitude envelopes \cite{marelli2010time}. Sound effects are usually produced by sound designers or foley artists by either recording the sounds on demand or sourcing and transforming the assets from pre-recorded sound libraries. However, with the increasing size and complexity of video games and interactive media, creating enough sound assets is time-consuming, especially in scenarios such as virtual reality (VR), where players may freely interact with elements of the virtual environment using haptic controllers \cite{proceduralavar}.

Alternatively to pre-recorded samples, sound effects can also be produced using sound synthesisers -- a method which is often called procedural audio \cite{farnell2010designing}. Typically, procedural audio models are handcrafted and built upon digital signal processing (DSP) algorithms running in real-time with parametric controls \cite{farnell2010designing}. Yet the process of building procedural audio models may be challenging for sound designers, and the resulting audio may also lack plausibility when compared to pre-recorded samples \cite{barahona2021specsingan, comunita2021neural}.
Differentiable digital signal processing (DDSP) \cite{engel2020ddsp} commonly refers to the concept of using DSP algorithms alongside deep learning. 
In the original DDSP paper, Engel et al.~\cite{engel2020ddsp} synthesise harmonic musical notes controlled by pitch and loudness using a harmonic plus noise synthesiser \cite{Serra1990}. Once trained, the resulting synthesiser is able to produce sounds with human-interpretable controls (e.g., pitch and loudness).

In the context of the synthesis of sound effects, and particularly in game audio, human-interpretable continuous controls are desirable, as the synthesiser could adapt its output to, for instance, in-game events (in the case of running in real-time) or to animations (running offline). DDSP-based models also benefit from requiring comparatively less data to train than other data-driven approaches \cite{engel2020ddsp}. Additionally, DDSP synthesisers have been demonstrated to be able to run in real-time \cite{ganis2021real}, offering the potential of being integrated into live scenarios such as video games.

\begin{centering}
\begin{figure*}[t]

\makebox[\textwidth][c]{\includegraphics[width=\textwidth]{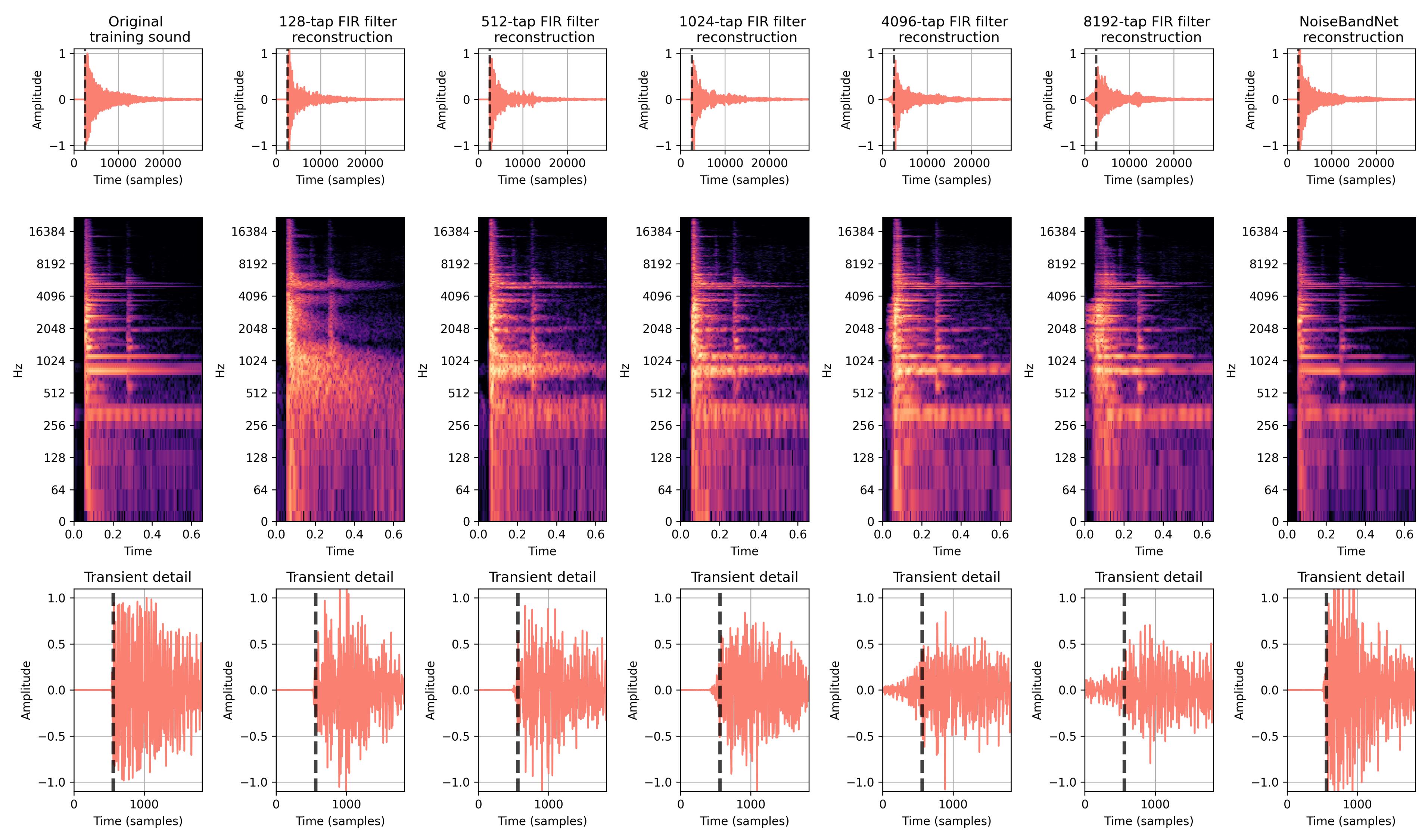}}%
\caption{Reconstruction task comparison between the DDSP time-varying FIR noise synthesiser and \modelname{}. The top row shows the waveform of the entire sound, the middle row its log-magnitude spectrogram and at the bottom a detail of the transient. The transient spot is annotated with a vertical dashed line in the first and third rows. The left column shows the original training sample: a short metal impact. The middle columns show the reconstruction of five different configurations of the DDSP time-varying FIR noise synthesiser with 128, 512, 1024, 4096 and 8192 taps respectively, all of them with a hop size of 32 samples. Observe its time and frequency trade-off: the frequency resolution increases with the number of taps at the same time the time resolution decreases, and vice-versa. The right column shows the \modelname{} reconstruction using 2048 filters and a synthesis window size of 32 samples, maintaining both good time and frequency resolution.}
\label{fig:ddsp_nbn_transient}
\end{figure*}
\end{centering}

Our end goals are to build a general-purpose DDSP synthesiser capable of producing a)~sounds with acceptable time and frequency resolution, and b)~audio of arbitrary length, just by providing conditioning vectors containing the desired parametric controls. The original DDSP synthesiser \cite{engel2020ddsp} relies on the premise that the audio it aims to model is harmonic, which is not the case for most sound effects. Very recently, \cite{hayes2022sinusoidal} proposed a method to estimate sinusoidal components using gradient descent, which has been a challenging task when using Fourier-based loss functions \cite{turian2020m}, opening the possibility of modelling inharmonic sinusoids using DDSP synthesisers. Sound effects, however, may contain noisy or very narrow-band elements that are difficult to model using sinusoidal partials \cite{marelli2010time}, plus the method of \cite{hayes2022sinusoidal} has yet to be applied to the context of sound effects. Another option could be to use a time-varying finite impulse response (FIR) filter as in the original DDSP subtractive noise synthesiser, but it suffers from a time and frequency trade-off, where, in order to obtain good frequency resolution, the number of taps in the FIR filter need to be relatively high, which in return smears the transients, and vice-versa. This phenomenon is depicted in Figure~\ref{fig:ddsp_nbn_transient}. As an alternative, and inspired by the work of \cite{marelli2010time} where they use multi-rate \fb{}s and sub-band processing to overcome the time and frequency trade-off of the inverse FFT method (very closely related to the DDSP FIR-noise synthesiser case), we explore the use of \fb{}s in this context, leading to a definition of a new architecture called \modelname{}. While we do not use sub-band processing in our work, we incorporate some of the ideas from \cite{marelli2010time} into a differentiable pipeline, linking human-interpretable control parameters to the output audio. We compare \modelname{} to the original DDSP noise synthesiser and establish a more suitable method to generate time-varying inharmonic sound effects of arbitrary length using DDSP synthesisers, with both good time and frequency resolution. Code and audio examples can be found at the project website.\footnote{\url{https://www.adrianbarahonarios.com/noisebandnet/}}

\section{Related Work}

There are multiple studies on the synthesis of sound effects using DSP techniques. From aeroacoustic \cite{selfridge2018creating} or footstep \cite{nordahl2010sound} sounds, to guidelines to choose appropriate synthesis methods \cite{farnell2010designing}, or efforts to bring models to the wider public \cite{bahadoran2018fxive}. Most current models, however, still lack plausibility when compared to pre-recorded samples \cite{barahona2021specsingan,comunita2021neural}.

Audio synthesis using deep learning, often called neural audio synthesis, can be seen as an alternative to pure DSP-driven synthesis. While this work has usually focused on speech or music signals, studies on the synthesis of sound effects exist. For instance, in \cite{barahona2020synthesising}, knocking sound effects are synthesised conditioned by emotions, and in \cite{comunita2021neural}, footstep sound effects are synthesised conditioned on surface materials. Other approaches focus on more general environmental sound categories, such as \cite{liu2021conditional, pascual2023full}. There has also been a growing interest in generating sound effects conditioned on natural language prompts, such as in \cite{kreuk2022audiogen, liu2023audioldm}. There is work addressing scarcity of training data when using neural audio synthesis too, especially relevant for sound effects. In \cite{greshler2021catch}, unconditional sound variations of arbitrary length are produced just by providing $\approx$20 seconds of data, and in \cite{barahona2021specsingan}, unconditional variations of short ($\approx$200-750 ms) one-shot sound effects are synthesised by training on a single audio example, which has also been applied to the task of data sonification \cite{pauletto2023sonifying}. More recently, \cite{andreu2022neural} generates novel sound effect variations of arbitrary length conditioned on mel-spectrograms, training on a small dataset.

DDSP architectures can be seen as a middle ground between models trained on large datasets over extended periods of time and architectures that generate data from a few examples, as they exploit inherent biases in DSP components such as filters or oscillators to facilitate the training, synthesis and control tasks. Apart from the original DDSP architecture \cite{engel2020ddsp}, other studies explore the use of waveshaping \cite{hayes2021neural}, wavetable \cite{shan2022differentiable}, or frequency modulation (FM) \cite{caspe2022ddx7} synthesis, all of them focusing on the modelling of harmonic or musical sounds. Other approaches, closer to sound effects, focus on the synthesis of rigid-body impacts by predicting the properties of resonant infinite impulse response (IIR) filters based on the object shape \cite{diaz2022rigid}, the synthesis of harmonic engine sounds \cite{lundberg2020data}, or footsteps using the original DDSP time-varying FIR noise synthesiser \cite{serrano2022neural}. In our case, we aim to provide a general-purpose method that, in principle, does not have any pre-conceptions about the target sound to be modelled, and that is able to render both wide and narrow spectral components of time-varying sounds.

Creatively controlling and conditioning deep learning audio models has also been studied previously. In \cite{nistal2020drumgan}, drum sounds are synthesised conditioned on timbral features. Other approaches such as \cite{okamoto2022onoma} drive the synthesis of environmental sounds using onomatopoeic words, perform real-time timbre transfer \cite{caillon2021rave}, use the latent space of a generative adversarial network (GAN) \cite{goodfellow2014generative} to condition a recurrent neural network (RNN) to synthesise sound textures \cite{wyse2022sound}, or generate music by pose sequences \cite{bisigraw}. Topics akin to the inverse of these control schemes have also been investigated, such as providing an audio example as input, and using a neural network approach to retrieve synthesiser control parameters \cite{esling2019flow}.

The use of \fb{}s in the context of sound texture synthesis was investigated in \cite{mcdermott2011sound}, where target sounds were decomposed into sub-bands using a cochlear \fb{}, and a set of statistics were extracted from their amplitude envelopes to build a synthesis model. In contrast to this, multi-rate \fb{}s have also been used to model time-varying environmental sounds with narrow spectral components, thus using the \fb{} structure itself to spectrally shape white noise \cite{marelli2010time}. They use sub-band processing to approximate the frequency response of a series of FIR filters that comprise a \fb{}, and multiply each of the noise bands resulting from filtering white noise though the \fb{} by a time-varying amplitude in order to match the target audio. 

\section{Method}\label{sect:method}
Our proposed method consists of using a deep learning model similar to the original DDSP architecture, conditioned on high-level audio controls to output $M$-channel time-varying amplitudes, and multiply them by the $M$ bands resulting from processing white noise through a \fb{}. From a high-level perspective, we first build a \fb{} comprised of adjacent FIR filters with narrow frequency responses that jointly cover an arbitrarily wide-ranging frequency spectrum. Then, in order to ease the computational burden of our approach, we precompute the filtering operation on a white noise instance with all the different filters of the \fb{}, ``baking'' those noise bands. Lastly, we use an architecture similar to the original DDSP paper to predict the time-varying amplitudes of each of the bands for a target dataset, conditioning it on high-level controls, and effectively linking control parameters to the output of the synthesiser. The final output is generated by summing all the bands together in the time-domain.

\subsection{Filterbank Design}

Since our method does not make assumptions about the frequency content of the sound to be modelled, and in order to allow for the synthesis of both narrow and broad frequency components, we need to design narrow bandpass filters, with the union of their combined frequency responses covering the whole frequency spectrum $[0, \text{Fs}/2]$, where $\text{Fs}$ is the sampling rate, which we set to $\text{Fs}=\SI{44.1}{\kilo\hertz}$. Thus, two adjacent bandpass filters will share one of their two band edges with each other to cover the totality of the frequency spectrum.

We start by deciding the number of filters $M$ that will comprise the \fb{} to obtain a good frequency resolution, which, based on pilot experiments, we set to $M=2048$ filters. Second, we decide how those filters are going to be distributed across the frequency spectrum. As in \cite{marelli2010time}, we emphasise the lower end of the frequency spectrum by covering those frequencies with more filters than at the higher end. Specifically, we use half of the filters (1024 in our case, $[1,...,1024]$) to cover the first quarter of the frequency spectrum $[0, \text{Fs}/8]$, distributing their center frequencies linearly in the interval. The other half of the filters (the other 1024 filters, $[1025,...,M]$) cover the remaining interval of the frequency spectrum, $[\text{Fs}/8, \text{Fs}/2]$, with their center frequencies spaced evenly on a logarithmic scale, thus increasing their bandwidth along it (i.e., filters at the higher end of the spectrum have a greater bandwidth than filters at the lower end). 

To implement the \fb{}, we use real FIR filters, designing them with the Kaiser window method with a transition width $\Delta_\omega$ of $20\%$ of the filter bandwidth:
\begin{equation}
\Delta_\omega = \frac{|\omega_{1}-\omega_{2}|}{\text{Fs}} \cdot 0.2
\end{equation}
where $\text{Fs}$ is the sampling rate, and $\omega_{1}$ and $\omega_{2}$ are the left and right band edges respectively. We use a stopband attenuation of $A_s=\SI{50}{\decibel}$. All the filters are bandpass except for the first one, which is a lowpass filter that covers the $[0, f_{\text{min}}]$ interval, with $f_{\text{min}}= \SI{20}{\hertz}$; and the last one, which is a highpass filter that covers the $[\omega_{1}, \text{Fs}/2]$ interval, were $\omega_{1}$ is the right band edge of the penultimate bandpass filter.

\begin{figure}
  \centering
    \includegraphics[width=0.5\textwidth]{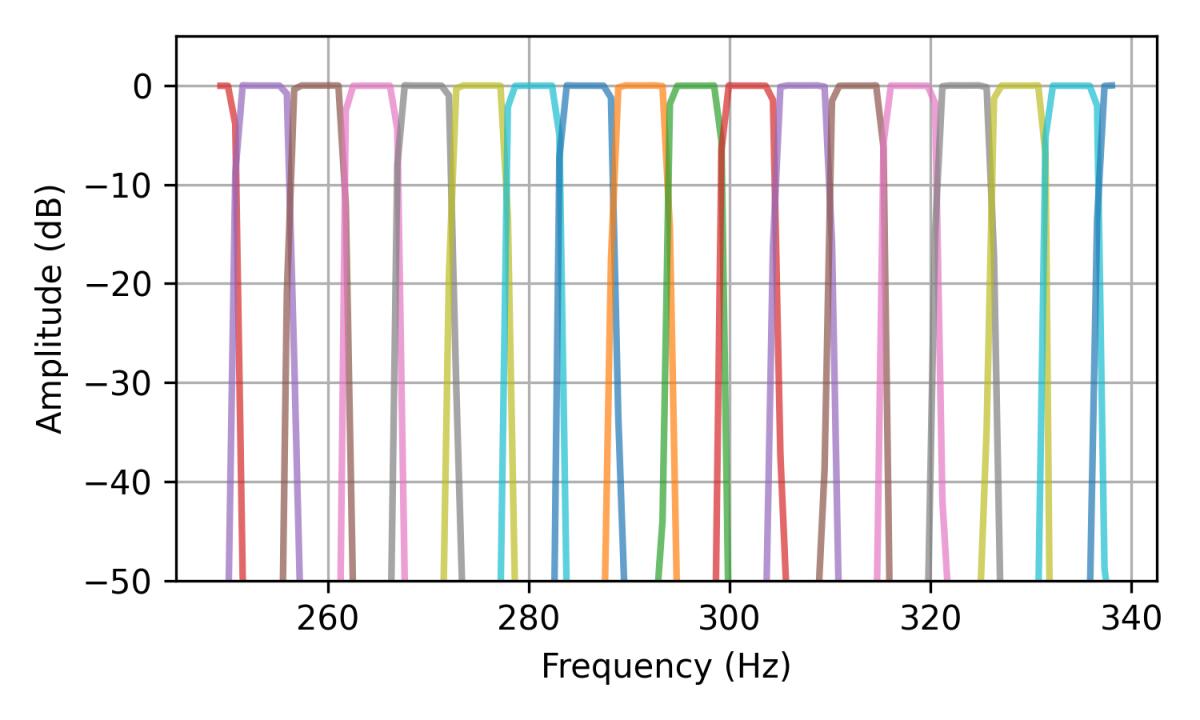}
  \caption{Detail of the frequency response of some of the filters employed in a 2048-filter \fb{}. Each of the filters is represented by a different colour.}
  \label{fig:filterbank_frequency_response}
\end{figure}

An example of the frequency response of some of the filters is depicted in Figure~{\ref{fig:filterbank_frequency_response}}. For reference, using the configuration described above, the longer FIR filter in the \fb{} has a total of \SI{120287} taps. The bandwidth of the linearly-distributed (bandpass) filters in the low end is $B_{(1,...,1024)}\approx$ \SI{5.4}{\hertz}, and the bandwidth of the last (highpass) filter is $B_{M}\approx$ \SI{30}{\hertz}. Once we build the \fb{}, we zero-pad all the filters' impulse responses to the next power of 2 of the length of the filter with more taps, so they all have the same length. Using the proposed configuration, this results in filters with \SI{131072} taps. The large length of these filters is due to their very narrow nature.

\subsection{Deterministic Loopable Noise Bands}\label{section:dlnb}

Considering we use many ($M=2048$) and long (\SI{131072}-tap) FIR filters in our system, generating the noise bands themselves (i.e., a convolving a noise instance with all the filters) is a computationally expensive operation, which can bottleneck both the training and inference of the model. This is especially true during training where, at each training step, the noise bands may need to be recalculated; or in longer sequences during inference (e.g., generating 120-seconds' worth of audio).

To ease the computational burden of our system, we follow the technique described in \cite{valimaki2018creating}, where they propose a method to extend stationary sounds such as airplane cabin noise, but applying it to the noise bands resulting from filtering a white noise instance with all the filters of the \fb{}. Our aim is to compute these noise bands only once and store (``bake'') them, removing the need of recomputing the operation each time the synthesiser generates an output.

More specifically, we use their proposed FFT convolution approach that leads to sounds that can be concatenated along their $x$-axis thanks to circular convolution. By the convolution theorem, it is known that convolution in the time-domain is equivalent to point-wise frequency-domain multiplication, which can be written as follows for the filtering of a white noise signal with an FIR filter \cite{valimaki2018creating}:
\begin{equation}
Y = R_{\text{noise}}R_{\text{filter}}e^{j(\theta_{\text{noise}}+\theta_{\text{filter}})},
\end{equation}
with $R$ and $\theta$ representing the magnitude and phase respectively resulting from the FFT. Since white noise ideally has a flat magnitude response, they set it to unity $R_{\text{noise}}=1$, and since the phase of the noise signal, $\theta_{\text{noise}}$, already randomises the phase of the operation ($\theta_{\text{noise}}+\theta_{\text{filter}}$), they replace it with a random phase, obtaining the final expression \cite{valimaki2018creating}:

\begin{equation}
Y = R_{\text{filter}}e^{j(\theta_{\text{random}})},
\label{eq:random_noise}
\end{equation}
where $\theta_{\text{random}}$ is formed by uniformly distributed random values drawn from a $[-\pi, \pi]$ interval and having its first and last values (DC and Nyquist frequencies, respectively) set to $0$ \cite{valimaki2018creating}. Since the FFT exhibits Hermitian symmetry for real-valued data, the values beyond the Nyquist frequency (the negative frequency values) are just the complex conjugate of the positive ones mirrored from the Nyquist frequency, excluding the Nyquist and DC elements. Finally, by taking the inverse FFT of Equation~\ref{eq:random_noise}, the ``loopable'' noise band is created due to the resulting circular convolution operation described above.

Using our proposed configuration, each of the noise bands have a length of \SI{131072} samples, corresponding to $\approx3$~seconds of audio at \SI{44.1}{\kilo\hertz}. We also enforce a deterministic behaviour by setting the same random seed each time a noise band is generated. This is done to 1)~maintain coherence each time noise bands are built (i.e., the noise bands used during training and inference will be identical) and 2)~being able to share the same noise band instances across multiple models, granting they have the same \fb{} configuration. Also, since the amplitude of each the resulting noise bands may be very small, due to the narrow portion of the frequency spectrum they focus on, we find the maximum amplitude value $A_{\text{max}}$ across all the noise bands that comprise the \fb{}, and divide all the bands by this $A_{\text{max}}$ value, effectively scaling their amplitudes up to what we found to be a reasonable level. While this leads to neither homogeneous amplitude distribution across bands, nor a normalised amplitude (i.e., in the range $[-1,1]$), when all the bands are summed together without further intervention, the scale of the individual noise bands will be handled by the time-varying amplitude predicted by the model (see Section~\ref{section:arch}).

\begin{figure}
  \centering
    \includegraphics[width=0.5\textwidth]{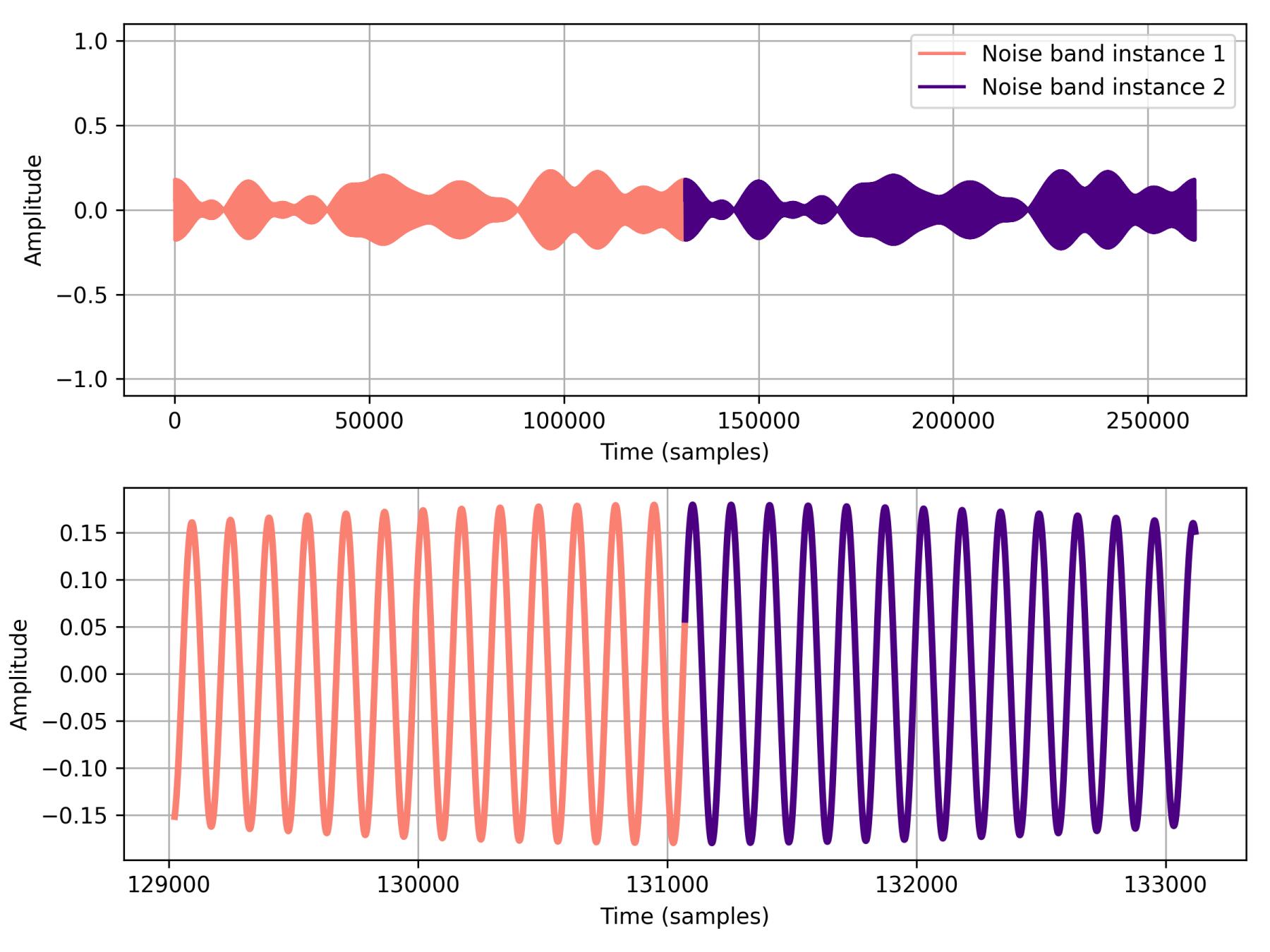}
  \caption{Loopable noise bands. Two instances of the same noise band are concatenated along their $x$-axis. The top figure shows the waveform of both noise bands, one after the other, each one with a distinctive colour. The bottom figure shows the detail of the point where the end of the first noise band instance meets the start of the second one. Notice how, thanks to circular convolution, the start and the end of the segments are ``joined up''.}
  \label{fig:noise_band_concatenation}
\end{figure}

Thus, by using the method proposed in \cite{valimaki2018creating}, we generate deterministic and loopable noise bands that only need to be computed once and can be extended arbitrarily in time by just concatenating them along their $x$-axis. An example of this is depicted in Figure~\ref{fig:noise_band_concatenation}, where two instances of the same noise band are concatenated, one after the other. Conceptually, each of those noise bands could be somewhat seen as a wavetable. A wavetable is defined as a block of memory (i.e., a ``table'') where a discretised signal is stored \cite{creasey2016audio} and, while they are usually employed to store a single period of a waveform, a loopable noise band may be regarded as a period -- as it can be looped -- of the portion of the frequency spectrum it captures.

\subsection{Architecture}\label{section:arch}

\begin{centering}
\begin{figure*}[t]
\makebox[\textwidth][c]{\includegraphics[width=\textwidth]{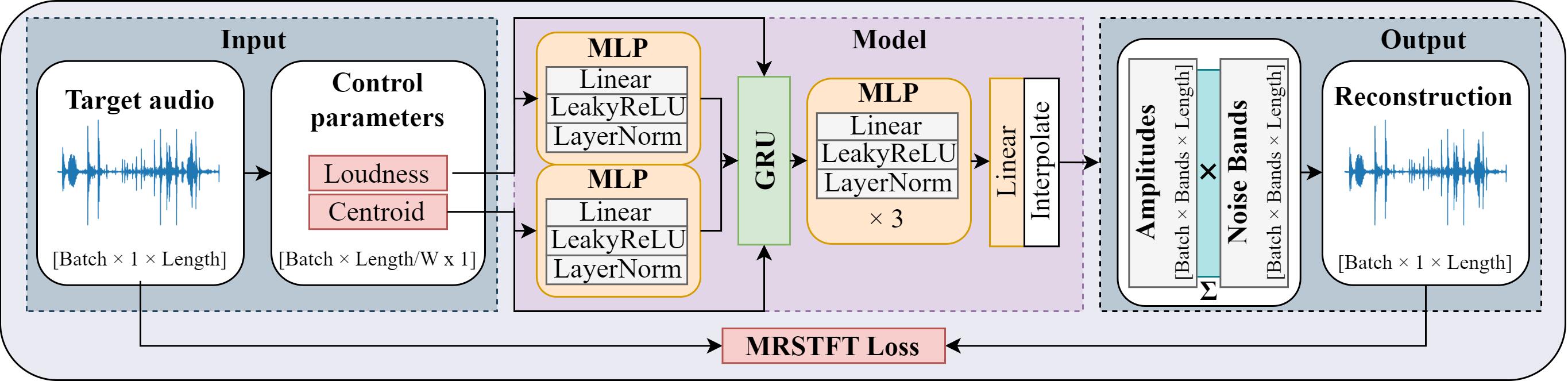}}%
\caption{Overview of the \modelname{} architecture and training process. In this case, loudness and spectral centroid features are extracted from the training audio and passed to the network, which predicts an $M$-band matrix of time-varying amplitudes at a $\text{Fs}$ sampling rate divided by a synthesis window size $W$. Depending on the control scheme, these features or control parameters may be different (e.g., only loudness or user-provided control parameters). The predicted amplitudes are upsampled using linear interpolation by a factor of $W$ to match the audio length, and multiplied by the $M$ noise bands. The output audio is generated by summing all the bands together. Finally, the model is optimised by comparing the resulting sound against the target audio using a multi-resolution STFT (MRSTFT) loss.}
\label{fig:nbn_arch}
\end{figure*}
\end{centering}

\modelname{}, depicted in Figure~\ref{fig:nbn_arch}, is built upon the original DDSP architecture \cite{engel2020ddsp}, but replacing their harmonic-plus-noise synthesiser with a \fb{} structure. As in DDSP, the internal sampling rate of the model is a fraction of the target dataset sampling rate $\text{Fs}$. To obtain a good time resolution, and as in \cite{marelli2010time}, we select a synthesis window size $W$ of 32 samples, granting the model an internal sampling rate of $\text{Fs}/W$, thus producing an amplitude value every $W$
samples. Greater $W$ values will lead to poorer time resolution but less computational burden, and vice-versa.

The inputs to the neural network component of \modelname{} are the control parameters, which in the Figure~\ref{fig:nbn_arch} are loudness and spectral centroid extracted from the training data. These control parameters may be different depending on the control scheme, such as only loudness, or other user-defined controls. Independently of the control scheme, and to synchronise the control parameters to the training data (i.e., to have a $1:1$ mapping between the control parameters and the samples in the target audio), originally the control parameters will have the same length as the dataset waveforms, interpolating them to this length if needed. Before passing the control parameter vectors to the network, we resample them according to  $\text{Fs}/W$, the internal sampling rate of our model.

Similar to \cite{engel2020ddsp}, the control vectors are passed through a time-distributed multi-layer perceptron (MLP) block (one per control parameter vector and in parallel, as depicted in Figure~\ref{fig:nbn_arch}) and a gated recurrent unit (GRU) \cite{bahdanau2014neural}. The output of the GRU is passed through a series of time-distributed MLP blocks leading to final time-distributed dense layer which outputs the the $M$-channel time-varying amplitudes (each one of them corresponding to each of the noise bands), with a sampling rate of $\text{Fs}/W$. As in \cite{engel2020ddsp}, to avoid negative amplitude values we scale the resulting amplitudes using a modified sigmoid activation function, in our case:

\begin{equation}
y = 2.0 \cdot \text{sigmoid}(x)^{log10}+10^{-18}
\end{equation}

Then, to bring them to audio rate $\text{Fs}$, we upsample these amplitudes by a factor of $W$ using linear interpolation. We then multiply the amplitudes by the noise bands in the time-domain, and sum them together to produce the final output audio. Unless stated otherwise, we model mono audio with a sampling rate $\text{Fs}$ of \SI{44.1}{\kilo\hertz}.

\subsection{Training and Inference}\label{sect:train}
We train the network on batches of audio chunks of length $L_{\text{chunk}}$. We concatenate all the training waveforms along the time dimension and select random chunks of length $L_{\text{chunk}}$ from them. This avoids the network memorising predicted amplitude values to the position of the training examples with respect of time, and so increases its generalisation capabilities when generating longer sequences (especially important when training with small datasets or one-shots). If the training dataset is comprised of a very short ($L_{\text{dataset}}<L_{\text{chunk}}$) training example, we simply repeat it along the $x$-axis until $L_{\text{dataset}} \geq L_{\text{chunk}}$.  As the control parameters have the same length as the audio, we select the same chunk and resample it to $\text{Fs}/W$ before passing them to the network.

Likewise, it is possible that the length of the training chunks $L_{\text{chunk}}$ may be smaller than the length of the noise bands $L_{\text{bands}}$. To prevent the network being exposed to portions of the noise bands that were never seen during training, we ``roll'' the noise bands along their $x$-axis at each training step, to a randomised integer shift drawn from a uniformly distributed random value in $[0, L_{\text{bands}}]$, achieving the use of a different, randomised portion of the noise bands at each training step. During training, we compare the output audio against the target audio using a multi-resolution STFT (MRSTFT) loss \cite{yamamoto2020parallel}, with the aim of reconstructing the input audio for the given control parameters. 

Once trained, the model needs only control parameter vectors of arbitrary length $L_{\text{control}}$ to produce an output of $L_{\text{control}}\cdot W$ length in samples. Due to the nature of the architecture used, this output is deterministic (i.e., the model produces the same output amplitudes for the same control parameter input). However, in practice, the output audio resulting from multiplying the noise bands by the same predicted amplitudes may be slightly different since, as described above, we randomise the start of the noise bands by a $[0, L_{\text{bands}}]$-shift, and their energy is not constant over their length (refer to  Figure~\ref{fig:noise_band_concatenation}, where the amplitude of the band fluctuates over time).

\section{Reconstruction}\label{section:reconstruction}

\begin{centering}

\begin{table*}[]
\centering

\begin{tabular}{lcccccccccccc}
\hline  \noalign{\vskip 0.01in}    

                   & \multicolumn{2}{c}{\textit{Footsteps}}           & \multicolumn{2}{c}{\textit{Thunderstorm}}        & \multicolumn{2}{c}{\textit{Pottery}}             & \multicolumn{2}{c}{\textit{Knocking}}            & \multicolumn{2}{c}{\textit{Metal}}                        & \multicolumn{2}{c}{Average values}               \\ 
\textbf{}          & \textit{\textbf{MRSTFT}} & \textit{\textbf{FAD}} & \textit{\textbf{MRSTFT}} & \textit{\textbf{FAD}} & \textit{\textbf{MRSTFT}} & \textit{\textbf{FAD}} & \textit{\textbf{MRSTFT}} & \textit{\textbf{FAD}} & \textit{\textbf{MRSTFT}} & \textit{\textbf{FAD}} & \textit{\textbf{MRSTFT}} & \textit{\textbf{FAD}} \\ \hline  \noalign{\vskip 0.02in}    
NoiseBandNet       & \textbf{1.14$\pm$0.01}   & \textbf{5.41}         & \textbf{1.24$\pm$0.01}   & \textbf{9.06}         & \textbf{1.22$\pm$0.04}   & 1.33                  & \textbf{1.10$\pm$0.01}   & \textbf{2.44}         & \textbf{1.15$\pm$0.01}   & \textbf{4.65}         & \textbf{1.17}            & \textbf{4.578}        \\ 
DDSP$_{\text{256 taps}}$  & 1.29$\pm$0.01            & 8.45                  & 1.44$\pm$0.01            & 10.08                 & 1.38$\pm$0.02            & 2.17                  & 1.32$\pm$0.01            & 8.68                  & 1.60$\pm$0.01            & 34.64                 & 1.40                     & 12.804                \\ 
DDSP$_{\text{512 taps}}$  & 1.26$\pm$0.01            & 9.22                  & 1.41$\pm$0.02            & 10.10                 & 1.37$\pm$0.02            & \textbf{1.22}         & 1.30$\pm$0.01            & 5.17                  & 1.46$\pm$0.01            & 28.75                 & 1.36                     & 10.89                 \\ 
DDSP$_{\text{1024 taps}}$ & 1.24$\pm$0.01            & 9.89                  & 1.42$\pm$0.01            & 10.33                 & 1.38$\pm$0.03            & 1.55                  & 1.29$\pm$0.01            & 5.35                  & 1.35$\pm$0.01            & 25.92                 & 1.34                     & 10.61                 \\ 
DDSP$_{\text{4096 taps}}$ & 1.22$\pm$0.02            & 7.02                  & 1.42$\pm$0.02            & 9.58                  & 1.39$\pm$0.04            & 2.06                  & 1.32$\pm$0.02            & 4.23                  & 1.27$\pm$0.01            & 15.53                 & 1.32                     & 7.69                  \\ \hline
\end{tabular}

\caption{MRSTFT loss ($\text{mean}\pm\text{sd}$) and FAD results for the reconstruction task on several sound categories comparing the DDSP FIR noise synthesiser using different configurations against \modelname{}. Lower values of MRSTFT loss and FAD are better (best performers highlighted in bold).}\label{table:resultstable}
\end{table*}
\end{centering}

First, we evaluate NoiseBandNet by comparing its reconstruction capabilities to different configurations of the original DDSP time-varying FIR noise synthesiser \cite{engel2020ddsp}. Their synthesiser produces an output by convolving white noise frame-by-frame with an FIR filter predicted by the network and then overlap-adding the frames. As in \cite{engel2020ddsp}, we do not model the FIR filters' impulse responses directly, but their magnitudes instead.

\subsection{Experiments}\label{section:experiments}

To evaluate the reconstruction capabilities of the systems, we select five sound effect categories relevant to video games which exhibit both broad and narrow spectral components and a wide range of amplitude envelopes \cite{farnell2010designing, barahona2020synthesising}:

\begin{itemize}
    \item Footsteps ($\approx$4.4~seconds): Footsteps on a metallic staircase.
    \item Thunderstorm ($\approx$14.0~seconds): Rain and close thunder sounds. 
    \item Pottery ($\approx$95.0~seconds): Breaking and scrapping pottery shards. 
    \item Knocking ($\approx$11.0~seconds): Knocking sound effects with different intensities and emotional intentions. 
    \item Metal ($\approx$19.0~seconds): Hitting and scrapping metal bars. 
\end{itemize}
We source all the training sound effects from the Freesound website \cite{font2013freesound}, except for the knocking sound effects, where we use an excerpt of the dataset provided by \cite{barahona2020synthesising}.

We choose loudness and spectral centroid to evaluate the reconstruction capabilities of the systems as they are related to the original DDSP loudness and pitch control vectors, but without the constraint of being harmonically-oriented. To extract the loudness and the spectral centroid, we use an FFT size of 128 and 512 respectively, both with 75\% overlap. For each sound category, we normalise each of the features to a $[0,1]$ range. Note this normalisation is dataset-dependent: we do not normalise the control parameters using their their full range (e.g., $[0, \text{Fs}/2]$ in the spectral centroid case), but using the maximum and minimum values computed for the feature across a particular dataset. This prevents feature values being localised to a small portion of the $[0, 1]$ interval for certain sounds (e.g., quieter sound categories would have most of their loudness values close to 0).

We train a \modelname{} model for each of the sound categories using a hidden size of 128 for all layers, a $M=2048$ band \fb{}, and 
a synthesis window $W$ of 32 samples, following the same design described in Section~\ref{sect:method}. We train all models for \SI{10000} epochs using a learning rate of 0.001, batch size of 16, an audio chunk size of \SI{65536} samples, Adam optimiser, and an MRSTFT loss \cite{yamamoto2020parallel} (with FFT sizes for the MRSTFT of $8192, 4096, 1024, 2048, 512, 128, 32$, 75\% overlap, and window lengths of the same size as the FFTs), employing the auraloss implementation \cite{steinmetz2020auraloss}.

Using an NVIDIA Tesla V100, the training process takes $\approx$45 min for all models, except for the pottery model, which takes $\approx$180 min. Once trained, the saved model weights have a size of $\approx$\SI{1.8}{\mega\byte}, with each model having a total of 464K parameters. During inference, the time required to synthesise a single batch signal with an output length of \SI{1322976} samples (around 30 seconds of audio at \SI{44.1}{\kilo\hertz}) is of $529.5\pm2.4$ ($\text{mean}\pm\text{sd}$) milliseconds on a consumer GPU (NVIDIA GTX 1060), and $13.4\pm0.3$ ($\text{mean}\pm\text{sd}$) seconds on a consumer CPU (AMD Ryzen 5 1600), measured on a 100-run test.

We evaluate \modelname{} resynthesis capabilities against four variants of the original DDSP model filtered noise synthesiser \cite{engel2020ddsp}, with a configuration of FIR filter taps of 256 (DDSP$_{\text{256 taps}}$), 512 (DDSP$_{\text{512 taps}}$), 1024 (DDSP$_{\text{1024 taps}}$) and 4096 (DDSP$_{\text{4096 taps}}$). We use a hop size of 32 samples for all of the models. While such hop size is small for some models compared to a more standard 75\% overlap, we use this value to 1)~compare \modelname{} and DDSP using a configuration that is as close as possible for all systems, and 2)~demonstrate that a smaller hop size does not necessarily improve the time resolution for the DDSP time-varying FIR noise synthesiser (refer to Figure~\ref{fig:ddsp_nbn_transient}). We use a hidden size of 128 for all of the models and a single input MLP per input feature, as in \modelname{} (see Figure~\ref{fig:nbn_arch}). We employ the DDSP noise synthesiser Pytorch implementation found in \cite{wu2022ddsp},\footnote{\url{https://github.com/YatingMusic/ddsp-singing-vocoders}} do not model reverb, and use the same training configuration and loss function as the \modelname{} models.

\subsection{Results}

We use two objective metrics to assess all five models' reconstruction fidelities. First, the MRSTFT loss described above, measured from 19 different values at training time as models converged near their final epochs, to compensate for small possible fluctuations occurring during training. Second, the Fr\'echet Audio Distance (FAD) \cite{kilgour2018fr}, a quality metric that correlates to human listeners better than spectral distances, using the implementation found in \footnote{\url{https://github.com/gudgud96/frechet-audio-distance}}. MRSTFT loss and FAD results are reported in Table~\ref{table:resultstable}.

A two-way ANOVA on the MRSTFT loss data with factors for model (five levels, of \modelname{} and the four variants of the original DDSP model) and sound effect (five levels of footsteps, thunderstorm, pottery sounds, knocking sound effects and metal sounds) reveals significant main effects of model ($F(4, 450) = 2128, p < .001$) and sound effect ($F(4, 450) = 1217, p < .001$) and a significant interaction ($F(16, 450) = 193, p < .001$), suggesting that the type of model drives differences in loss, so does the type of sound effect, and that certain combinations of model and sound effect lead to either particularly low or high loss values.

An analysis of multiple pairwise comparisons (Tukey's Honest Significant Difference method) was conducted to investigate which pairings of groups differ. It was found that \modelname{} significantly outperforms DDSP$_{\text{256 taps}}$ ($\textrm{mean diff} = 0.234,\ p < .001$), DDSP$_{\text{512 taps}}$ ($\textrm{mean diff} = 0.189,\ p < .001$), DDSP$_{\text{1024 taps}}$ ($\textrm{mean diff} = 0.164,\ p < .001$) and DDSP$_{\text{4096 taps}}$ ($\textrm{mean diff} = 0.150,\ p < .001$). Thus, \modelname{} obtains significantly better MRSTFT reconstruction values compared to the variants of the original DDSP noise synthesiser. In terms of the DDSP model variants' performance across the different sound effect categories, they were most effective for footsteps, followed by knocking, pottery, and thunder, with relatively small differences in performance between variants. This was in contrast to the metal category, where DDSP variants displayed relatively large differences in performance.

The FAD results follow a very similar pattern to those for MRSTFT, but we note the exception that DDSP$_{\text{512 taps}}$ performs better than \modelname{} in terms of FAD for the pottery sound effect category. This discrepancy may be caused by the small size of our datasets, which negatively affects the accuracy of the metric \cite{kilgour2018fr}.

\section{Creative Uses}\label{section:creative-uses}

As a second evaluatory perspective on \modelname{}, in this section we highlight some potential creative uses to which \modelname{} can be put.

\subsection{Amplitude Randomisation}\label{sect:randomisation}

\begin{figure}
  \centering
    \includegraphics[width=0.5\textwidth]{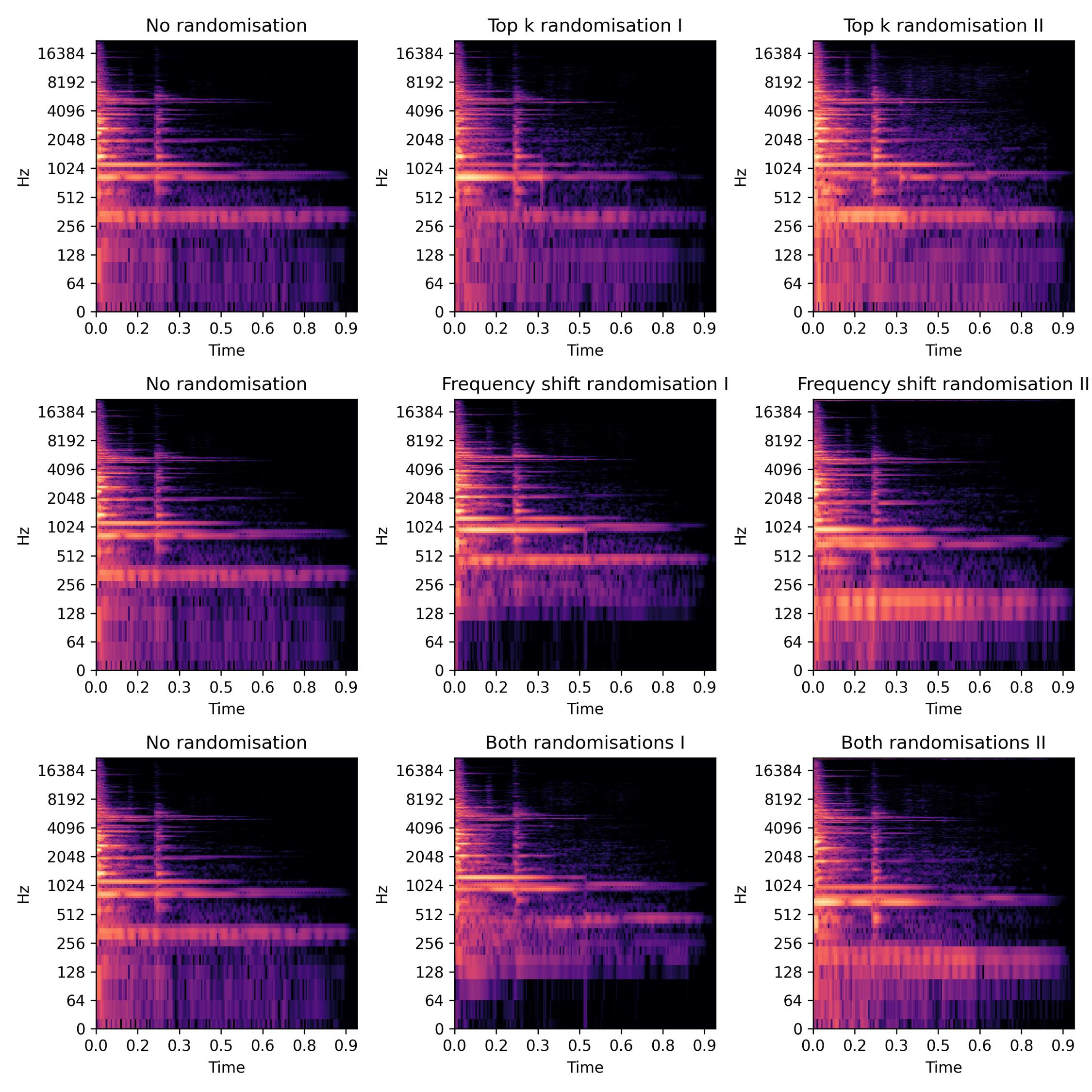}
  \caption{Log-magnitude spectrograms from the result of the different randomisation schemes. The left column represents a non-randomised (just reconstructed) sound: a metal impact. The second and third columns show two examples of the resulting randomised sound. In the first row we employ the top $k$ randomisation scheme using $L_{\text{frame}}=430$ (3 frames), $k=100$ and a randomised multiplier in a $[0.0, 1.0]$ range. The second row depicts the frequency shift randomisation scheme with $L_{\text{frame}}=645$  (2 frames), $f_{\text{init}}=30$ and  $f_{\text{shift}}=3$. The third row shows both randomisation combined, using $L_{\text{frame}}=645$  (2 frames), $k=100$, a $[0.0, 1.0]$ multiplier, $f_{\text{init}}=30$ and  $f_{\text{shift}}=3$.}
  \label{fig:nbn_randomisation}
\end{figure}

Given that \modelname{} uses DSP components at its core to produce audio, we can exploit their inherent biases to further alter the output signal. As outlined in Section~\ref{sect:train}, the output amplitudes of the model using the proposed architecture is deterministic. Here, as an example, we present two strategies to generate variations from the predicted time-varying amplitudes. This may be especially relevant to game audio, where it is common to use multiple audio clips to sound design the same in-game interaction in order to avoid repetition \cite{zdanowicz2019game}.

First, we propose to randomise the top $k$ amplitudes $k_{\text{amp}}$ within a desired frame length $L_{\text{frame}}$ (i.e., we randomise the output amplitudes each $L_{\text{frame}}$ amplitude values). To achieve this, first we select the frame length $L_{\text{frame}}$ and split the output amplitudes to non-overlapping frames of that length. Note we split these frames before performing the linear interpolation operation that upsample the amplitude values to audio rate. Then, we find the desired top amplitudes $k_{\text{amp}}$ on each frame by summing the amplitude values for each of the bands on that frame across the time-axis and selecting the greater $k$ values. After that we apply a randomised amplitude modifier in an user-defined range of $[mult_{\text{min}}, mult_{\text{max}}]$ by multiplying the amplitude values on that frame by it, scaling them up or down. Since all amplitudes still need to be interpolated to audio rate, the transition between their values is smoothed. We also found that, if we compute a different amplitude randomisation for the same amplitude output we can generate stereo sequences by panning them left and right as the resulting signal, in combination to the variation introduced by the band shift explained in Section~\ref{sect:train}, will be slightly different for relatively small $[mult_{\text{min}}, mult_{\text{max}}]$ values.

\begin{centering}
\begin{figure*}[t]

\makebox[\textwidth][c]{\includegraphics[width=\textwidth]{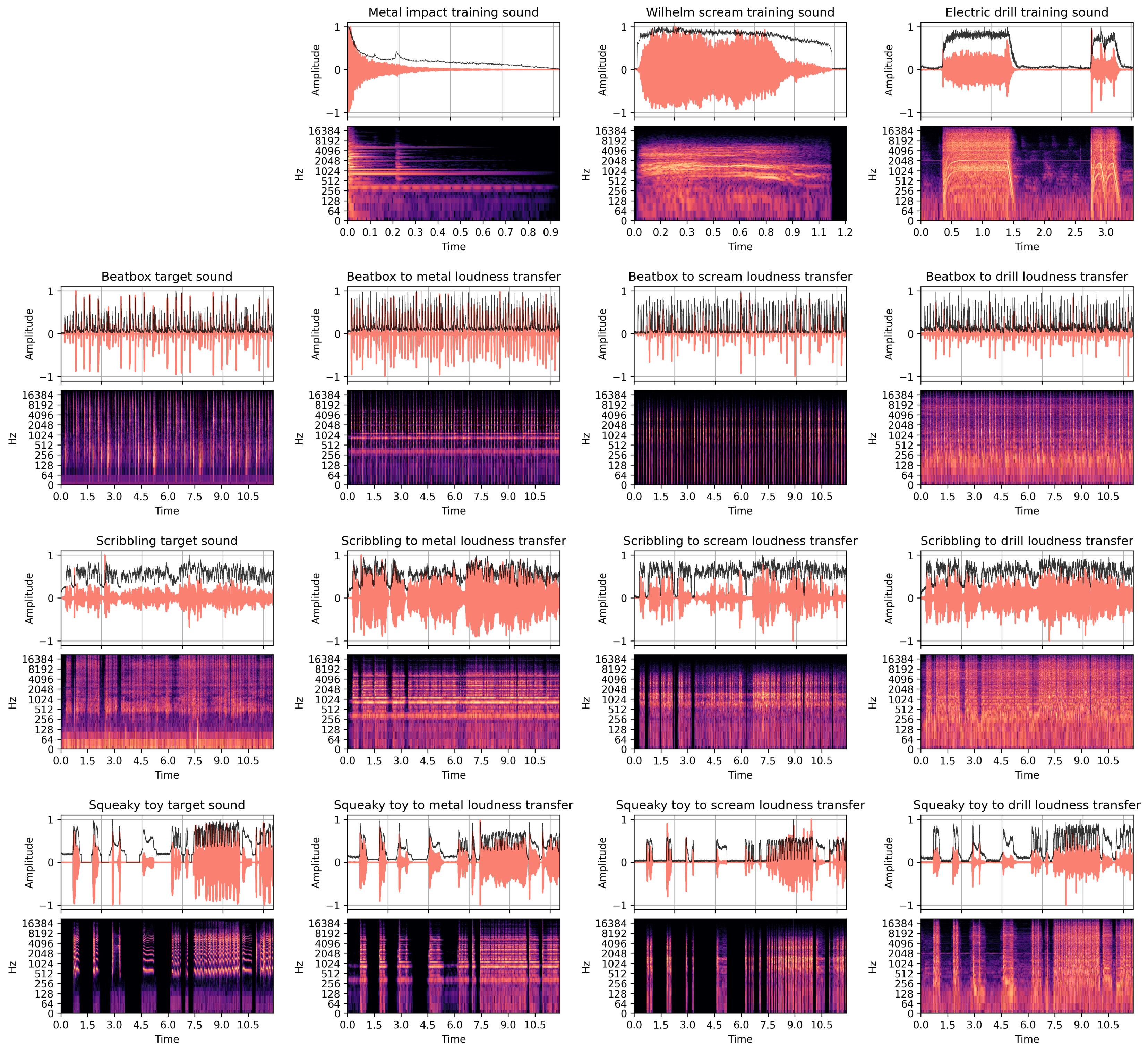}}%
\caption{Waveforms (top) and log-magnitude spectrograms (bottom) pairs resulting from the loudness transfer experiments. The extracted loudness of each sound is represented in black, superimposed on the waveforms in a $[0,1]$ range. The first row depicts the three sounds used to train each of the models: a metal impact, the Wilhelm scream. and an electrical drill sound effect. The first column contains the sounds used for transferring their loudness envelopes. Starting from the second row, the second, third, and fourth columns contain the loudness transfer results for the different sound combinations.}
\label{fig:loudness_transfer}
\end{figure*}
\end{centering}

Second, we propose another strategy to perform frame-wise pitch-shift on the output amplitudes. Again, we select a desired frame length $L_{\text{frame}}$ and split the output amplitudes to non-overlapping frames of it before the linear interpolation operation. Within that frame, we ``roll'' all the amplitude values to a randomised integer value  $[-f_{\text{shift}}, f_{\text{shift}}]$ on their band-axis, effectively transposing the bands from one amplitude to another. We take into account the previous $f_{\text{shift}}$ values to compute the current shift, implementing a process somewhat similar to a random walk. We also allow for an initial frequency shift $f_{\text{init}}$ that rolls all the amplitude values in a randomised $[-f_{\text{init}}, f_{\text{init}}]$ range, effectively transposing the entire sound. Likewise, the subsequent linear interpolation operation to audio rate will provide a relative smooth transition between shifts.

An example of both schemes is depicted in Figure~\ref{fig:nbn_randomisation}, showing the top $k$ randomisation in the first row, the frequency shift randomisation in the second row and both randomisation schemes applied together in the third row.

\subsection{Loudness Transfer}\label{subs:loudness_transfer}

In \cite{engel2020ddsp} they were capable of performing timbre transfer (i.e., transferring the pitch and loudness of an incoming audio to the instrument the model is trained on) using just 13 minutes of expressive solo violin performances. However, unlike harmonic sounds that are constrained by a discretised and well-defined pitch, in Section~\ref{section:experiments} we use spectral centroid as an alternative control vector for inharmonic sounds (such as in most sound effects), thus introducing a higher degree of freedom to the control parameters. Considering the deterministic nature of the output amplitudes from the model, and since obtaining enough expressive data to represent all possible loudness and spectral centroid and interactions may be challenging in the context of sound effects, here employ a control scheme that only relies in one of the features: loudness. Our aim is to transfer the relative loudness envelope of one sound to another, training our network in the latter and using the extracted loudness envelope of the former during inference. This is possible due to loudness being mathematically defined (i.e., it can be extracted programmatically) and normalised to a $[0,1]$ range relative to the training and inference data, as described in Section~\ref{section:experiments}, thus covering the full loudness range regardless of the training data. 

To demonstrate the loudness transfer capabilities of the model we follow the same training procedure than in Section~\ref{section:experiments}, but using loudness as the only control parameter. We train three different models on the following short sounds: a metal impact ($\approx1.0$ second), the Wilhelm scream ($\approx1.2$ seconds), and an electric drill sound ($\approx3.4$ seconds). Due to having a single control parameter and therefore a single input MLP, the trained models are slightly smaller than the ones trained on two control parameters. More specifically, they have a total of $\approx414K$ parameters (as opposed to 464K) and their weighs a size of $\approx$\SI{1.6}{\mega\byte} (as opposed to $\approx$\SI{1.8}{\mega\byte}). For reference, the MRSTFT reconstruction loss for the different models is of $1.04\pm0.01$ for the metal impact model, $1.09\pm0.01$ for the Wilhelm scream model and $1.17\pm0.01$ for the drill model using the same objective as in Section~\ref{section:experiments}.

We then choose another three sounds to transfer their loudness envelope to all the trained models: a rhythmic beatbox sound effect, scribbling using a pencil onto paper sounds, and a squeaky toy sound effect. We collect all the training and inference sound effects from the Freesound website \cite{font2013freesound}. The loudness transfer is performed by simply extracting the loudness from the target sounds (beatbox, scribbling and drill in our examples) computing it using the same procedure described in Section~\ref{section:experiments} (including normalising it to a  $[0,1]$ range), and using the resulting vector (interpolated accordingly to the internal $\text{Fs}/W$ sampling rate) as the input to the trained models (metal impact, Wilhelm scream and electric drill). We choose a $2^{19}$ sample length excerpt from the target sounds ($\approx12$s at $\SI{44.1}{\kilo\hertz}$) and apply the operation described above. 

While both the loudness of the target and trained models sounds are normalised in a $[0,1]$ range, it may occur that most (or the most perceptually relevant) of their values are contained on an specific interval, and outliers distort it. To solve this potential issue, and to allow for a finer finer control over the output of the model, we apply an user-defined scale modifier to the loudness values. In our experiments, we applied the following modifiers to the input loudness values of the sounds depicted in Figure~\ref{fig:loudness_transfer}: metal impact ($+0.1$ on beatbox, $-0.1$ on scribbling, $-0.1$ on squeaky toy), Wilhelm scream ($+0.3$ on beatbox, $+0.15$ on scribbling, no modification on squeaky toy), electric drill ($+0.25$ on beatbox, no modification in scribbling, $-0.1$ on squeaky toy).

The result from the experiments is depicted in Figure~\ref{fig:loudness_transfer}. It can be discerned how the target loudness envelope, depicted in the first column is successfully transferred to the trained models, depicted in the second, third and fourth columns, starting from the second row. It is also noticeable how the frequency content of the resulting transferred sounds is time-varying, changing over time depending on the input control parameter.

\subsection{Training and Synthesis using User-Defined Control Parameters}

Since loudness (or spectral centroid) curves may be challenging to control and interpret from an user-perspective, or may not be adequate for the potential use-cases of a particular model, here we explore training on user-provided control parameters, taking inspiration from the Wwise audio middleware Real-Time Parameter Controls (RTPCs)\footnote{\url{https://www.audiokinetic.com/en/library/edge/?source=SDK&id=concept_rtpc.html}}. RTPCs can be used to attach in-game parameters (e.g., the speed of a car) to sound properties (e.g., the pitch of the engine), linking game events to sound control curves. In our scenario, by anticipating the creative use of the synthesiser, an user (e.g., a sound designer) may directly draw the desired control curves used during training and, once trained, use them to control the output of the model. 

\begin{figure}
  \centering
    \includegraphics[width=0.5\textwidth]{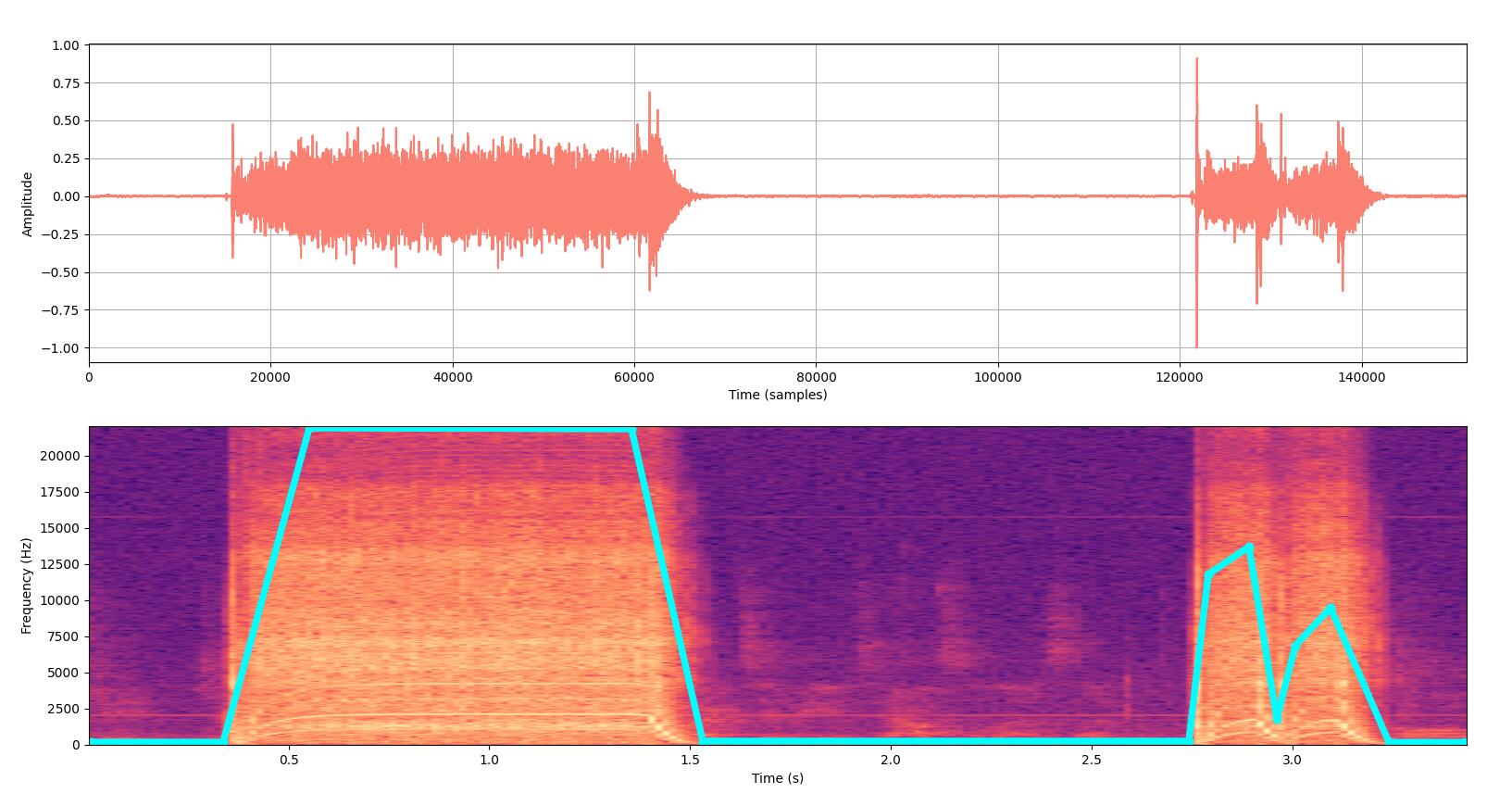}
  \caption{Graphical user interface used to manually label the data. The image depicts the waveform (top) and the magnitude spectrogram (bottom) of an electric drill sound. The cyan line on top of the spectrogram is the hand-annotated control curve.}
  \label{fig:drill_ui_label}
\end{figure}

To this end, we design a graphical user interface to manually label the data based on potential control curves. The tool is depicted in Figure~\ref{fig:drill_ui_label}, containing a waveform of the sound to be modelled on top and its spectrogram at the bottom. By clicking on top of the spectrogram, a user can manually draw the control curve. Once drawn, this curve is normalised to $[0,1]$, in preparation for use with the model. We choose the same three sounds used in Section~\ref{subs:loudness_transfer} and envisage a user wishing to the following aspects with their control curves: for metal impact, the curve might control impact force; for the Wilhelm scream, the curve might control scream intensity; for the electric drill, the curve might control drill power. We then train the three models with those hand-drawn control curves as their single input control parameter. For reference, in this case the MRSTFT reconstruction loss for the different models is of $1.05\pm0.01$ for the metal impact model, $1.01\pm0.01$ for the Wilhelm scream model and $1.23\pm0.01$ for the drill model, using the same objective and configuration as in Section~\ref{section:experiments}.

\begin{centering}
\begin{figure*}[t]

\makebox[\textwidth][c]{\includegraphics[width=\textwidth]{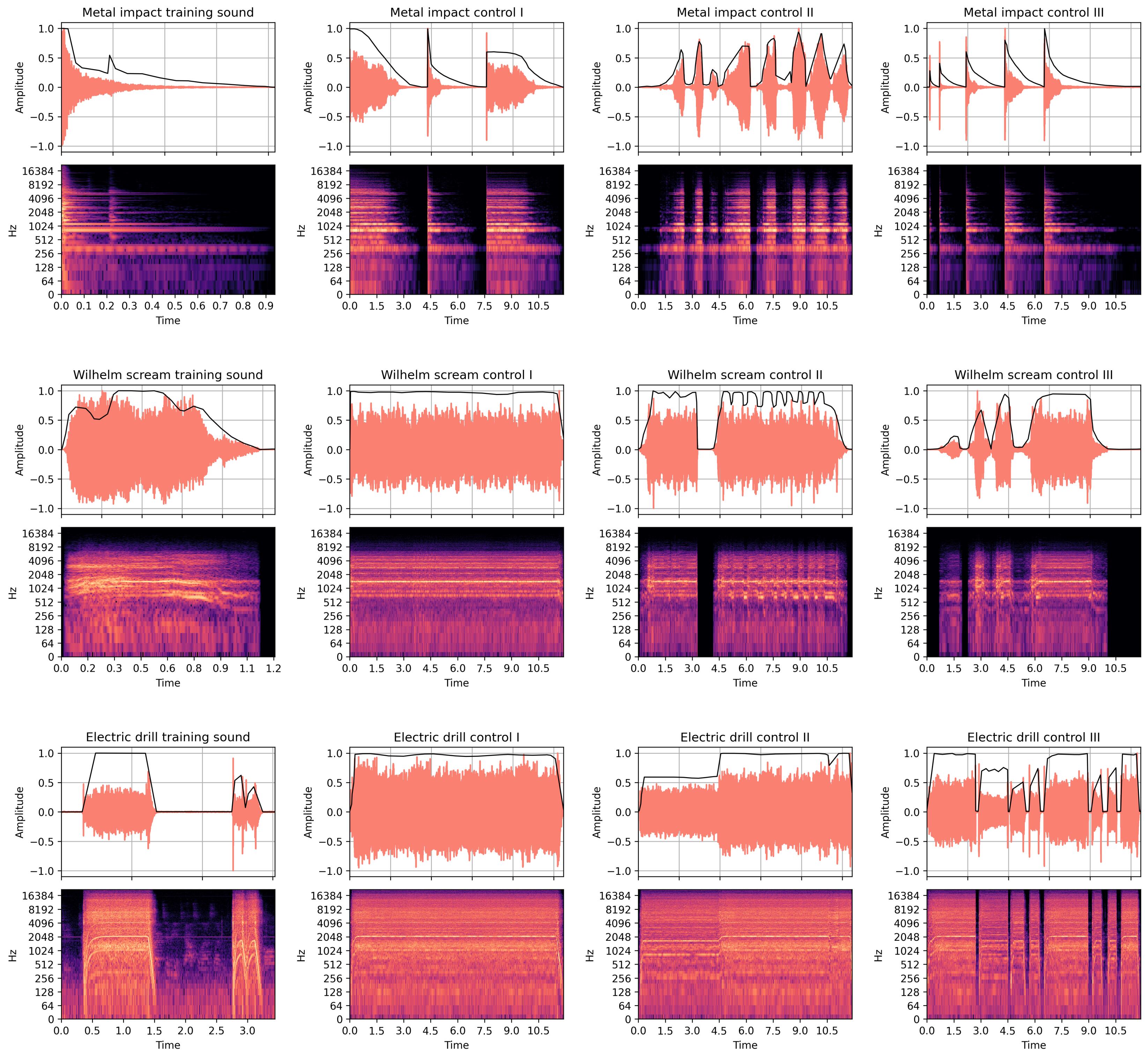}}%
\caption{Waveforms (top) and log-magnitude spectrograms (bottom) pairs resulting from the training on user-defined control experiments. The training sounds are depicted in the first column, with their user-defined training control parameters represented in black superimposed on the waveforms in a $[0,1]$ range. The second, third, and fourth columns contain the sounds generated by using user-defined inference curves for the three models, each one in a different row. The user-defined inference control curves are also represented in black on top of the waveforms in a $[0,1]$ range.}
\label{fig:control_manual}
\end{figure*}
\end{centering}

To control the synthesiser, we provide a corresponding UI tool for drawing the inference control parameters. The control tool functions exactly as the tool depicted in Figure~\ref{fig:drill_ui_label}, but now the user has a blank canvas to draw their desired control curve for driving the synthesiser. We draw three hand-crafted curves per model with a length of $2^{14}$ each (second, third, and fourth columns of Figure~\ref{fig:control_manual}). Since we upsample the output amplitudes to audio rate interpolating them by a factor defined by the synthesis window size $W=32$, the output signal length is of $2^{14}\cdot32 = $ \SI{524288} samples or $\approx$12~seconds at $\SI{44.1}{\kilo\hertz}$. The results of the experiments are depicted in Figure~\ref{fig:control_manual}. The first column contains the original sounds along their user-defined control curves used during training, represented in black on top of the waveforms. The subsequent columns are the synthesised sounds resulting from using the new user-defined inference curves, depicted also in black on top of their waveforms. For each sound category, it can be seen that the resulting audio is broadly consistent with the intended control curve.

\section{Discussion}

How high-fidelity sound effects can be generated 1) automatically or 2) with dynamic or creative control where necessary or desired -- all without compromising the quality or plausibility of the output audio -- is a topic of interest to the fields of sound design and game audio \cite{farnell2010designing, zdanowicz2019game}, psychoacoustics \cite{mcdermott2011sound}, and extended reality \cite{proceduralavar}. It is a topic for which DDSP methods have shown promise in recent years, at least where assumptions hold regarding the harmonicity of the sounds being modeled \cite{engel2020ddsp}. Harmonic sounds represent only a portion of sound effects, however, and so it remained an open problem how to model and then synthesise \emph{arbitrary sounds} with acceptable time and frequency resolution, and of arbitrary length.

The contribution of this paper is to address the modelling and synthesis of arbitrary sounds, tackling both the problem of reconstruction fidelity (see Section~\ref{section:reconstruction}) and exploring some of the creative uses (see Section~\ref{section:creative-uses}). Our solution is encapsulated in a model called \modelname{}, an architecture capable of synthesising continuous sound effects conditioned on high-level parametric controls with consistently good time and frequency resolution. We propose the use of \fb{}s to shape white noise, establishing a suitable approach towards modelling non-musical or inharmonic sound effects using DDSP synthesisers. \modelname{} is also lightweight and can be trained on very limited data ($\approx$1~second of audio), as shown in our experiments. We also highlight the potential creative uses of the architecture by generating sound variations, performing loudness transfer, and training and synthesising audio with user-defined control parameters.

We evaluated \modelname{} against four configurations of the original DDSP filtered noise synthesiser \cite{engel2020ddsp}, and found that \modelname{} significantly outperforms all DDSP variants on the task of resynthesising sounds from different categories, for nine out of ten (sound category, evaluation metric)-combinations -- the exception being for the metric of FAD on pottery sounds. In addition to overall better reconstruction capabilities compared to the original DDSP noise synthesiser, our proposed \fb{} is not constrained by having its frequency response distributed linearly, such as in the case of a time-varying FIR filter. Thus, both the number of filters and their distribution across the frequency spectrum is flexible and can be altered to accommodate other use cases. 

Taking inspiration from current game audio workflows, we also outlined the creative possibilities of \modelname{} through a series of experiments, providing examples of amplitude randomisation, automatic loudness transfer and training models using user-defined controls. The code employed to generate those sounds alongside the audio examples described throughout the paper can be found in the accompanying material at the project website.\footnote{\url{https://www.adrianbarahonarios.com/noisebandnet/}}

\subsection{Limitations and Future Work}

While we used a \fb{} configuration with a higher frequency resolution on the low end, broadly inspired by \cite{marelli2010time}, which provided satisfactory results on pilot experiments, the design could be further improved by considering auditory perception, for instance increasing the emphasis between $500$ and ${\SI{4000}\hertz}$, where the sensitivity of frequency changes to pure tones is higher \cite{oxenham2018we}. Apart from the effect of the number of filters and their distribution on synthesis quality, we also plan to explore the use of alternative loss functions, such as the differentiable joint time-frequency scattering (JTFS), used recently in the context of audio classification with promising results \cite{muradeli2022differentiable}.

Since the proposed noise band structure is not tied to the network architecture itself, for future work we aim to use noise bands with other approaches. For instance, we could replace the architecture with a more lightweight temporal convolution network (TCN) \cite{bai2018empirical}, which has been successfully employed to model audio effects \cite{steinmetz2021efficient} and in differentiable FM synthesisers \cite{caspe2022ddx7}. Another option may be using adversarial training \cite{goodfellow2014generative} or a variational autoencoder (VAE) \cite{kingma2013auto, rezende2014stochastic}, which opens up the possibility of non-deterministic behaviour. Additionally, \modelname{} could be applied to harmonic and musical signals, replacing the original DDSP noise synthesiser, or potentially in combination with it when sounds contain inharmonic partials (e.g., training using an approach derived from \cite{hayes2022sinusoidal}). 

Despite the saved model weights being small in size ($\approx$\SI{1.8}{\mega\byte} and $\approx$\SI{1.6}{\mega\byte} for two and one control parameters, respectively), the size of the noise bands is large ($\approx$\SI{1}{\giga\byte} on disk for our configuration), due to their the number and length. However, as described in Section~\ref{section:dlnb}, thanks to the deterministic nature of the noise bands when using the same \fb{} configuration, a single instance can be used across multiple models, thus only needing to create a single set of them. Nonetheless, to further optimise the model size and alleviate the computational burden involved in multiplying the output amplitudes from the model with the noise bands in the time-domain, we plan to investigate the use of neural audio codecs such as Encodec \cite{defossez2022high}, and multi-rate \fb{}s and sub-band processing as in \cite{marelli2010time}. As we reported in Section~\ref{section:experiments}, the offline generation on a consumer GPU is fast ($\approx$529.5~milliseconds to generate $30$~seconds of audio), but it is much slower on CPU ($\approx$13.4~seconds to generate the same length). While more research needs to be conducted to address these points, we hypothesise that a combination of architectural changes (such as the use of TCNs), a more efficient auditory-informed \fb{} configuration, and the use of neural audio codecs and sub-band processing as described above may result in faster generation, which is especially relevant for real-time synthesis in the context of game audio.

Regarding the automatic extraction of control parameters from the audio, above (Section~\ref{section:experiments}) we use loudness and spectral centroid, computed using DSP methods. Other approaches, such as \cite{lundberg2020data}, develop highly engineered solutions to extract control parameters from the target audio, such as engine RPM in their case. It is desirable, however, to accommodate a wider range of sounds and use cases. While a first approach could be the use of sound event detection to extract similar sounding clips from longer signals in data-scarce scenarios, such as in \cite{wang2020few}, achieving the potential granularity required to successfully label continuous data (e.g., drill power in our examples) may be challenging. Another direction, inspired by the recent proliferation of text-to-audio models such as \cite{kreuk2022audiogen, liu2023audioldm}, could be the generation short audio clips catered towards being controlled by a model such as \modelname{}. For instance, a text-to-audio model could be prompted to generate a drill sound effect with a linearly increasing drill power, and the output audio could be used as the input to \modelname{} alongside a linearly increasing control parameter vector going from $[0,1]$ (minimum and maximum drill power values), granting the text-to-audio model successfully renders a sound with those properties.

We acknowledge that while we present three different experiments exploring the creative uses of the architecture, these could be expanded and evaluated in a user study. Future work will comprise carrying out a study with audio experts to evaluate the creative possibilities of the model, and the plausibility of the synthesised sounds. The study will also inform the amount and type of data needed to satisfactorily accomplish a control task, and the feasibility of training with multiple user-defined control parameters (e.g., a weather audio model with both ``rain and thunder intensity'' control curves). Since \modelname{} uses DSP components (time-varying amplitudes applied to filters) that audio experts are familiar with, we also plan to evaluate and expand the randomisation schemes outlined in Section~\ref{sect:randomisation}. Ultimately, we aim to understand how the model introduced in this paper could affect the workflows of sound designers and, more generally, audio experts in years to come when using controllable neural audio synthesisers in the context of game audio.

\section*{Acknowledgments}
This work was supported by the EPSRC Centre for Doctoral Training in Intelligent Games \& Game Intelligence (IGGI) [EP/L015846/1] and Sony Interactive Entertainment Europe.

\bibliographystyle{IEEEtran}
\bibliography{bib}

\begin{thebibliography}{10}
\providecommand{\url}[1]{#1}
\csname url@samestyle\endcsname
\providecommand{\newblock}{\relax}
\providecommand{\bibinfo}[2]{#2}
\providecommand{\BIBentrySTDinterwordspacing}{\spaceskip=0pt\relax}
\providecommand{\BIBentryALTinterwordstretchfactor}{4}
\providecommand{\BIBentryALTinterwordspacing}{\spaceskip=\fontdimen2\font plus
\BIBentryALTinterwordstretchfactor\fontdimen3\font minus
  \fontdimen4\font\relax}
\providecommand{\BIBforeignlanguage}[2]{{%
\expandafter\ifx\csname l@#1\endcsname\relax
\typeout{** WARNING: IEEEtran.bst: No hyphenation pattern has been}%
\typeout{** loaded for the language `#1'. Using the pattern for}%
\typeout{** the default language instead.}%
\else
\language=\csname l@#1\endcsname
\fi
#2}}
\providecommand{\BIBdecl}{\relax}
\BIBdecl

\bibitem{hausman2015modern}
C.~Hausman, F.~Messere, and P.~Benoit, \emph{{Modern Radio and Audio
  Production: Programming and Performance}}.\hskip 1em plus 0.5em minus
  0.4em\relax Cengage Learning, 2015.

\bibitem{marelli2010time}
D.~Marelli, M.~Aramaki, R.~Kronland-Martinet, and C.~Verron, ``{Time-Frequency
  Synthesis of Noisy Sounds With Narrow Spectral Components},'' \emph{IEEE
  transactions on audio, speech, and language processing}, vol.~18, no.~8, pp.
  1929--1940, 2010.

\bibitem{proceduralavar}
``{Post-Keynote Panel: Procedural Sound Synthesis for AR/VR},''
  \url{https://www.youtube.com/live/9ngwhfF0FhA?feature=share&t=6329/},
  accessed: 2023-07-10.

\bibitem{farnell2010designing}
A.~Farnell, \emph{{Designing Sound}}.\hskip 1em plus 0.5em minus 0.4em\relax
  Mit Press, 2010.

\bibitem{barahona2021specsingan}
A.~Barahona-R{\'\i}os and T.~Collins, ``{SpecSinGAN: Sound Effect Variation
  Synthesis Using Single-Image GANs},'' in \emph{19th Sound and Music Computing
  Conference, Saint-Étienne, France}, 2022.

\bibitem{comunita2021neural}
M.~Comunit{\`a}, H.~Phan, and J.~D. Reiss, ``{Neural Synthesis of Footsteps
  Sound Effects with Generative Adversarial Networks},'' in \emph{Audio
  Engineering Society Convention 152}.\hskip 1em plus 0.5em minus 0.4em\relax
  Audio Engineering Society, 2022.

\bibitem{engel2020ddsp}
J.~Engel, L.~Hantrakul, C.~Gu, and A.~Roberts, ``{DDSP: Differentiable Digital
  Signal Processing},'' \emph{arXiv preprint arXiv:2001.04643}, 2020.

\bibitem{Serra1990}
\BIBentryALTinterwordspacing
``{Spectral Modeling Synthesis: A Sound Analysis/Synthesis System Based on a
  Deterministic Plus Stochastic Decomposition},'' \emph{Computer Music
  Journal}, vol.~14, no.~4, p.~12, 1990. [Online]. Available:
  \url{http://www.jstor.org/stable/3680788?origin=crossref}
\BIBentrySTDinterwordspacing

\bibitem{ganis2021real}
F.~Ganis, E.~F. Knudesn, S.~V. Lyster, R.~Otterbein, D.~S{\"u}dholt, and
  C.~Erkut, ``{Real-time Timbre Transfer and Sound Synthesis using DDSP},''
  \emph{arXiv preprint arXiv:2103.07220}, 2021.

\bibitem{hayes2022sinusoidal}
B.~Hayes, C.~Saitis, and G.~Fazekas, ``{Sinusoidal Frequency Estimation by
  Gradient Descent},'' in \emph{ICASSP 2023-2023 IEEE International Conference
  on Acoustics, Speech and Signal Processing (ICASSP)}.\hskip 1em plus 0.5em
  minus 0.4em\relax IEEE, 2023.

\bibitem{turian2020m}
\BIBentryALTinterwordspacing
J.~Turian and M.~Henry, ``{I’m Sorry for Your Loss: Spectrally-Based Audio
  Distances Are Bad at Pitch},'' in \emph{''I Can't Believe It's Not Better!''
  NeurIPS 2020 workshop}, 2020. [Online]. Available:
  \url{https://openreview.net/forum?id=Z4UwGkTRTes}
\BIBentrySTDinterwordspacing

\bibitem{selfridge2018creating}
R.~Selfridge, D.~Moffat, E.~J. Avital, and J.~D. Reiss, ``{Creating Real-Time
  Aeroacoustic Sound Effects Using Physically Informed Models},'' \emph{Journal
  of the Audio Engineering Society}, 2018.

\bibitem{nordahl2010sound}
R.~Nordahl, S.~Serafin, and L.~Turchet, ``{Sound Synthesis and Evaluation of
  Interactive Footsteps for Virtual Reality Applications},'' in \emph{2010 IEEE
  Virtual Reality Conference (VR)}.\hskip 1em plus 0.5em minus 0.4em\relax
  IEEE, 2010, pp. 147--153.

\bibitem{bahadoran2018fxive}
P.~Bahadoran, A.~Benito, T.~Vassallo, and J.~D. Reiss, ``{Fxive: A Web Platform
  for Procedural Sound Synthesis},'' in \emph{AES Convention 144}.\hskip 1em
  plus 0.5em minus 0.4em\relax Audio Engineering Society, 2018.

\bibitem{barahona2020synthesising}
A.~Barahona-R{\'\i}os and S.~Pauletto, ``{Synthesising Knocking Sound Effects
  Using Conditional WaveGAN},'' in \emph{17th Sound and Music Computing
  Conference, Torino, Italy}, 2020.

\bibitem{liu2021conditional}
X.~Liu, T.~Iqbal, J.~Zhao, Q.~Huang, M.~D. Plumbley, and W.~Wang,
  ``{Conditional Sound Generation Using Neural Discrete Time-Frequency
  Representation Learning},'' in \emph{2021 IEEE 31st International Workshop on
  Machine Learning for Signal Processing (MLSP)}.\hskip 1em plus 0.5em minus
  0.4em\relax IEEE, 2021.

\bibitem{pascual2023full}
S.~Pascual, G.~Bhattacharya, C.~Yeh, J.~Pons, and J.~Serr{\`a}, ``{Full-Band
  General Audio Synthesis With Score-Based Diffusion},'' in \emph{ICASSP
  2023-2023 IEEE International Conference on Acoustics, Speech and Signal
  Processing (ICASSP)}.\hskip 1em plus 0.5em minus 0.4em\relax IEEE, 2023.

\bibitem{kreuk2022audiogen}
F.~Kreuk, G.~Synnaeve, A.~Polyak, U.~Singer, A.~D{\'e}fossez, J.~Copet,
  D.~Parikh, Y.~Taigman, and Y.~Adi, ``{AudioGen: Textually Guided Audio
  Generation},'' \emph{arXiv preprint arXiv:2209.15352}, 2022.

\bibitem{liu2023audioldm}
H.~Liu, Z.~Chen, Y.~Yuan, X.~Mei, X.~Liu, D.~Mandic, W.~Wang, and M.~D.
  Plumbley, ``{AudioLDM: Text-To-Audio Generation With Latent Diffusion
  Models},'' \emph{arXiv preprint arXiv:2301.12503}, 2023.

\bibitem{greshler2021catch}
G.~Greshler, T.~Shaham, and T.~Michaeli, ``{Catch-A-Waveform: Learning to
  Generate Audio from a Single Short Example},'' \emph{{Advances in Neural
  Information Processing Systems}}, vol.~34, pp. 20\,916--20\,928, 2021.

\bibitem{pauletto2023sonifying}
S.~Pauletto, A.~Barahona-R{\'\i}os, V.~Madaghiele, and Y.~Seznec, ``{Sonifying
  Energy Consumption Using SpecSinGAN},'' in \emph{20th Sound and Music
  Computing Conference, Stockholm, Sweden}, 2023.

\bibitem{andreu2022neural}
S.~Andreu and M.~V. Aylagas, ``{Neural Synthesis of Sound Effects Using
  Flow-Based Deep Generative Models},'' in \emph{Proceedings of the AAAI
  Conference on Artificial Intelligence and Interactive Digital Entertainment},
  vol.~18, no.~1, 2022, pp. 2--9.

\bibitem{hayes2021neural}
B.~Hayes, C.~Saitis, and G.~Fazekas, ``{Neural Waveshaping Synthesis},''
  \emph{arXiv preprint arXiv:2107.05050}, 2021.

\bibitem{shan2022differentiable}
S.~Shan, L.~Hantrakul, J.~Chen, M.~Avent, and D.~Trevelyan, ``{Differentiable
  Wavetable Synthesis},'' in \emph{IEEE International Conference on Acoustics,
  Speech and Signal Processing (ICASSP)}, 2022.

\bibitem{caspe2022ddx7}
F.~Caspe, A.~McPherson, and M.~Sandler, ``{DDX7: Differentiable FM Synthesis of
  Musical Instrument Sounds},'' \emph{arXiv preprint arXiv:2208.06169}, 2022.

\bibitem{diaz2022rigid}
R.~Diaz, B.~Hayes, C.~Saitis, G.~Fazekas, and M.~Sandler, ``{Rigid-Body Sound
  Synthesis with Differentiable Modal Resonators},'' in \emph{ICASSP 2023-2023
  IEEE International Conference on Acoustics, Speech and Signal Processing
  (ICASSP)}.\hskip 1em plus 0.5em minus 0.4em\relax IEEE, 2023.

\bibitem{lundberg2020data}
A.~Lundberg, ``{Data-Driven Procedural Audio: Procedural Engine Sounds Using
  Neural Audio Synthesis},'' Master's thesis, KTH School of Electrical
  Engineering and Computer Science (EECS), 2020.

\bibitem{serrano2022neural}
D.~Serrano, ``{A Neural Analysis--Synthesis Approach to Learning Procedural
  Audio Models},'' Master's thesis, New Jersey Institute of Technology,
  Department of Computer Science, 2022.

\bibitem{nistal2020drumgan}
J.~Nistal, S.~Lattner, and G.~Richard, ``{DrumGAN: Synthesis of Drum Sounds
  With Timbral Feature Conditioning Using Generative Adversarial Networks},''
  \emph{arXiv preprint arXiv:2008.12073}, 2020.

\bibitem{okamoto2022onoma}
Y.~Okamoto, K.~Imoto, S.~Takamichi, R.~Yamanishi, T.~Fukumori, Y.~Yamashita
  \emph{et~al.}, ``{Onoma-To-Wave: Environmental Sound Synthesis From
  Onomatopoeic Words},'' \emph{APSIPA Transactions on Signal and Information
  Processing}, 2022.

\bibitem{caillon2021rave}
A.~Caillon and P.~Esling, ``{RAVE: A Variational Autoencoder for Fast and
  High-Quality Neural Audio Synthesis},'' \emph{arXiv preprint
  arXiv:2111.05011}, 2021.

\bibitem{goodfellow2014generative}
I.~Goodfellow, J.~Pouget-Abadie, M.~Mirza, B.~Xu, D.~Warde-Farley, S.~Ozair,
  A.~Courville, and Y.~Bengio, ``{Generative Adversarial Nets},''
  \emph{{Advances in Neural Information Processing Systems}}, vol.~27, 2014.

\bibitem{wyse2022sound}
L.~Wyse, P.~Kamath, and C.~Gupta, ``{Sound Model Factory: An Integrated System
  Architecture for Generative Audio Modelling},'' in \emph{International
  Conference on Computational Intelligence in Music, Sound, Art and Design
  (Part of EvoStar)}.\hskip 1em plus 0.5em minus 0.4em\relax Springer, 2022,
  pp. 308--322.

\bibitem{bisigraw}
D.~Bisig and K.~Tatar, ``{Raw Music from Free Movements: Early Experiments in
  Using Machine Learning to Create Raw Audio from Dance Movements},'' \emph{2nd
  Conference on AI Music Creativity}, 2021.

\bibitem{esling2019flow}
P.~Esling, N.~Masuda, A.~Bardet, R.~Despres, and A.~Chemla-Romeu-Santos,
  ``{Flow Synthesizer: Universal Audio Synthesizer Control With Normalizing
  Flows},'' \emph{Applied Sciences}, vol.~10, no.~1, p. 302, 2019.

\bibitem{mcdermott2011sound}
J.~H. McDermott and E.~P. Simoncelli, ``{Sound Texture Perception via
  Statistics of the Auditory Periphery: Evidence From Sound Synthesis},''
  \emph{Neuron}, vol.~71, no.~5, pp. 926--940, 2011.

\bibitem{valimaki2018creating}
V.~V{\"a}lim{\"a}ki, J.~R{\"a}m{\"o}, and F.~Esqueda, ``{Creating Endless
  Sounds},'' in \emph{Proc. 21st Int. Conf. Digital Audio Effects (DAFx-18),
  Aveiro, Portugal}, 2018, pp. 32--39.

\bibitem{creasey2016audio}
D.~Creasey, \emph{{Audio Processes: Musical Analysis, Modification, Synthesis,
  and Control}}.\hskip 1em plus 0.5em minus 0.4em\relax Routledge, 2016.

\bibitem{bahdanau2014neural}
D.~Bahdanau, K.~H. Cho, and Y.~Bengio, ``{Neural Machine Translation by Jointly
  Learning to Align and Translate},'' in \emph{3rd International Conference on
  Learning Representations, ICLR}, 2015.

\bibitem{yamamoto2020parallel}
R.~Yamamoto, E.~Song, and J.-M. Kim, ``{Parallel Wavegan: A Fast Waveform
  Generation Model Based on Generative Adversarial Networks With
  Multi-Resolution Spectrogram},'' in \emph{ICASSP 2020-2020 IEEE International
  Conference on Acoustics, Speech and Signal Processing (ICASSP)}.\hskip 1em
  plus 0.5em minus 0.4em\relax IEEE, 2020, pp. 6199--6203.

\bibitem{font2013freesound}
F.~Font, G.~Roma, and X.~Serra, ``{Freesound Technical Demo},'' in
  \emph{Proceedings of the 21st ACM international conference on Multimedia},
  2013, pp. 411--412.

\bibitem{steinmetz2020auraloss}
C.~J. Steinmetz and J.~D. Reiss, ``{auraloss: Audio-Focused Loss Functions in
  PyTorch},'' in \emph{Digital Music Research Network One-day Workshop}, 2020.

\bibitem{wu2022ddsp}
D.-Y. Wu, W.-Y. Hsiao, F.-R. Yang, O.~D. Friedman, W.~Jackson, Y.-W. Liu, Y.-H.
  Yang \emph{et~al.}, ``{DDSP-Based Singing Vocoders: A New Subtractive-Based
  Synthesizer and a Comprehensive Evaluation},'' in \emph{Ismir 2022 Hybrid
  Conference}, 2022.

\bibitem{kilgour2018fr}
K.~Kilgour, M.~Zuluaga, D.~Roblek, and M.~Sharifi, ``{Fr\'echet Audio Distance:
  A Metric for Evaluating Music Enhancement Algorithms},'' \emph{arXiv preprint
  arXiv:1812.08466}, 2018.

\bibitem{zdanowicz2019game}
G.~Zdanowicz and S.~Bambrick, \emph{{The Game Audio Strategy Guide: A Practical
  Course}}.\hskip 1em plus 0.5em minus 0.4em\relax Focal Press, 2019.

\bibitem{oxenham2018we}
A.~J. Oxenham, ``{How We Hear: The Perception and Neural Coding of Sound},''
  \emph{Annual review of psychology}, vol.~69, pp. 27--50, 2018.

\bibitem{muradeli2022differentiable}
J.~Muradeli, C.~Vahidi, C.~Wang, H.~Han, V.~Lostanlen, M.~Lagrange, and
  G.~Fazekas, ``{Differentiable Time-Frequency Scattering On GPU},'' in
  \emph{Digital Audio Effects Conference (DAFx)}, 2022.

\bibitem{bai2018empirical}
S.~Bai, J.~Z. Kolter, and V.~Koltun, ``{An Empirical Evaluation of Generic
  Convolutional and Recurrent Networks for Sequence Modeling},'' \emph{arXiv
  preprint arXiv:1803.01271}, 2018.

\bibitem{steinmetz2021efficient}
C.~J. Steinmetz and J.~D. Reiss, ``{Efficient Neural Networks for Real--Time
  Modeling of Analog Dynamic Range Compression},'' in \emph{Audio Engineering
  Society Convention 152}.\hskip 1em plus 0.5em minus 0.4em\relax Audio
  Engineering Society, 2022.

\bibitem{kingma2013auto}
D.~P. Kingma and M.~Welling, ``{Auto-Encoding Variational Bayes},'' \emph{arXiv
  preprint arXiv:1312.6114}, 2013.

\bibitem{rezende2014stochastic}
D.~J. Rezende, S.~Mohamed, and D.~Wierstra, ``{Stochastic Backpropagation and
  Approximate Inference in Deep Generative Models},'' in \emph{International
  conference on machine learning}.\hskip 1em plus 0.5em minus 0.4em\relax PMLR,
  2014, pp. 1278--1286.

\bibitem{defossez2022high}
A.~D{\'e}fossez, J.~Copet, G.~Synnaeve, and Y.~Adi, ``{High Fidelity Neural
  Audio Compression},'' \emph{arXiv preprint arXiv:2210.13438}, 2022.

\bibitem{wang2020few}
Y.~Wang, J.~Salamon, N.~J. Bryan, and J.~P. Bello, ``{Few-Shot Sound Event
  Detection},'' in \emph{ICASSP 2020-2020 IEEE International Conference on
  Acoustics, Speech and Signal Processing (ICASSP)}.\hskip 1em plus 0.5em minus
  0.4em\relax IEEE, 2020, pp. 81--85.

\end{thebibliography}

\end{document}